\documentclass[12pt]{article}
\usepackage{amsmath}
\usepackage[hmarginratio=1:1,top=25mm,left=15mm,columnsep=18pt]{geometry}
\usepackage{amssymb}
\usepackage{paralist}
\usepackage{graphicx}
\usepackage{titling}
\usepackage[font=footnotesize,labelfont=bf]{caption}
\usepackage{float}
\usepackage{titlesec}
\usepackage{titling}
\usepackage{lipsum}
\usepackage{lineno}
\usepackage{tango}
\usepackage[]{hyperref}
\usepackage{authblk}
\hypersetup{
	bookmarksnumbered=true,     
	allcolors=blue,      
	colorlinks=true,            
	pdfpagemode=UseOutlines,    % this is the option you were lookin for
	linkcolor=chameleon4,
	citecolor=skyblue2,
	urlcolor=skyblue2
}
\let\OLDthebibliography\thebibliography
\renewcommand\thebibliography[1]{
	\OLDthebibliography{#1}
	\setlength{\parskip}{0pt}
	\setlength{\itemsep}{0pt plus 0.3ex}
}

\titleformat{\section}
{\normalfont\large\bfseries}{\thesection}{1em}{}

\title{\LARGE{\textbf {Stable memory with unstable synapses}}}
\date{}
\posttitle{\par\end{center}}

\author[1,2]{Lee Susman}
\author[2,3]{Naama Brenner}
\author[2,4]{Omri Barak}
\affil[1]{Interdisciplinary program in Applied Mathematics, Technion Israel Institute of Technology, Haifa, Israel}
\affil[2]{{Network Biology Research Laboratories, Technion Israel Institute of Technology, Haifa, Israel}}
\affil[3]{Faculty of Chemical Engineering, Technion Israel Institute of Technology, Haifa, Israel}
\affil[4]{Rappaport Faculty of Medicine, Technion Israel Institute of Technology, Haifa, Israel}
\begin{document}
%\large	
%\twocolumn[
\begin{@twocolumnfalse}
	\maketitle
	\begin{abstract} 
		What is the physiological basis of long-term memory? 
		The prevailing view in neuroscience attributes changes in synaptic efficacy to memory acquisition. 
		This view implies that stable memories correspond to stable connectivity patterns. 
		However, an increasing body of experimental evidence points to significant, activity-independent dynamics in synaptic strengths. 
		Motivated by these observations, we explore the possibility of memory storage within a global component of network connectivity, while individual connections fluctuate. 
		We find a simple and general principle, stemming from stability arguments, that links eigenvalues in the complex plane to memories. 
		Specifically, imaginary-coded memories are more resilient to noise and homeostatic plasticity than their real-coded counterparts. 
		Memory representations are stored as time-varying attractors in neural state-space and support associative retrieval of learned information. 
		Our results suggest a link between the properties of learning rules and those of network-level memory representations, and point at measurable signatures to be sought in experimental data. 
		{\vspace{1cm}}
	\end{abstract}
\end{@twocolumnfalse}
%]
%\noindent\makebox[\linewidth]{\rule{\paperwidth}{0.4pt}}

%\begin{multicols}{2}
\section*{}
%\linenumbers     % add line numbers!

\vspace{-0.9cm}
The ability to form and retain memories of past experience is fundamental to behavior, supporting adaptable responses and future planning \cite{Clayton03}. 
These internal representations persist over extended durations and may be reactivated by appropriate retrieval cues \cite{KahanaFoundations12}. 
Currently, it is widely accepted that synaptic connections between neurons play a central role in the physiological basis of long-term memory storage \cite{PooEngramReview16} (see \cite{Tsien13,Titley17} for other possibilities). 
The process of learning, on its part, is understood as stimulus-driven neural activity sculpting network architecture, i.e. Hebbian plasticity \cite{PooSTDP2001}.

If an internal memory-representation is stable over time, then one could assume that some properties of its underlying neural implementations also exhibit invariance over this period. 
However, at the level of single synapses, no such robustness exists (reviewed in \cite{MongilloVolatility2017,ClopathVariance2017,ChambersReview2017,ZivTenacity17}). Over the past decade, several studies, both ex vivo \cite{ZivTenacity2009,Yasumatsu2008} and in vivo \cite{LoewensteinSpines2015}, suggest that synapses undergo significant spontaneous changes. 
These fluctuations persist even in the absence of neural activity, with magnitude estimated to be as large as that of directed, Hebbian, plasticity \cite{DvorkinRemodeling2016}.
 
How, then, can memory traces remain stable over time? 
Various studies have proposed candidate invariant features, at different levels of organization of neural networks.
%\cite{MongilloVolatility2017}. 
For single synapses, invariance may be implemented in a sub-set of the largest spines \cite{LoewensteinSpines2015,Gan09}. 
Invariance may, instead, only emerge at the level of the connection between neurons, typically comprising several synapses. 
This allows individual synapses to fluctuate, under the constraint of stable overall connection strength between two cells \cite{Koester05,Fares09}.
Higher up the organizational hierarchy, invariant features may manifest only at the network level. 
This would allow individual connections to fluctuate, provided that some network properties remain invariant \cite{ChambersReview2017}.

In this work, we show that 
%stability of network dynamics imply that 
the combination of activity-independent synaptic fluctuations with known homeostatic mechanisms suggests a natural segregation of synaptic modifications at the network level. 
Such a segregation is supported by general arguments of system stability.
Specifically, fluctuations erode information encoded in the real part of the eigenvalues of synaptic connectivity, while sparing the imaginary-coded information. 
Such imaginary-coded memories correspond to anti-symmetric synaptic modifications, that can arise from Spike Timing Dependent Plasticity (STDP), which has a temporally asymmetric profile \cite{PooSTDP98}. In this scenario single synapses exhibit ongoing fluctuations, whereas invariance emerges as a network-level property. 

We investigate this concept by showing how different homeostatic plasticity mechanisms degrade real- or imaginary-coded memories, and how STDP can store transient inputs as imaginary-coded memories. 
We then show the implications of such memories - the learned representations give rise to stable oscillatory trajectories of network activity.
These memory states can be viewed as the time-varying analogs of stable fixed points in the classic Hopfield model \cite{Hopfield82}.
%the latter comprise a better account of network dynamics during retrieval, in light of an immense body of work pointing at population-level oscillatory activity [REFS].
After being learned and embedded in a component of connectivity, memory items may be transiently retrieved by supplying an associative recall cue. 

Our results suggest a principle by which memory can be learned and retained in a stable manner despite significant ongoing synaptic fluctuations. 
The implications of such a mechanism to experimental data are discussed both in terms of measured neural activity and in terms of synaptic plasticity during learning as opposed to at rest.

\section*{Results}

Our model is based on a standard framework of firing-rate neural networks \cite{DayanAbbottBook}. 
It consists of $N$ recurrently connected neurons, with $W_{ij}$ the synaptic connection strength from neuron $j$ to $i$. 
Each neuron $i$ transforms its input $x_i$ into firing rate via a nonlinearity $\phi\left(x_i\right)$, where the state vector $\mathbf{x} = \left(x_1 ~ x_2 ~ \cdots ~ x_N \right)^T$ evolves as
\begin{equation}
\dot{\bf{x}} = -\mathbf{x} +\mathbf{W}\phi\left(\mathbf{x}\right) + \mathbf{b}\left(t\right), \label{eq:xdot}
\end{equation}
and $\mathbf{b}$ is an external input.
Here and below we denote by $\phi(\mathbf{x})$ the vector obtained by applying $\phi$ to each coordinate of $\bf{x}$.

%This framework, while abstract, has proven useful in capturing qualitative aspects of network-level phenomena \cite{ManteNature03,BurakFiete2009Grid,Russo2018,Barak2017}.

\begin{figure}[H]	
	
	\includegraphics[width=\textwidth]{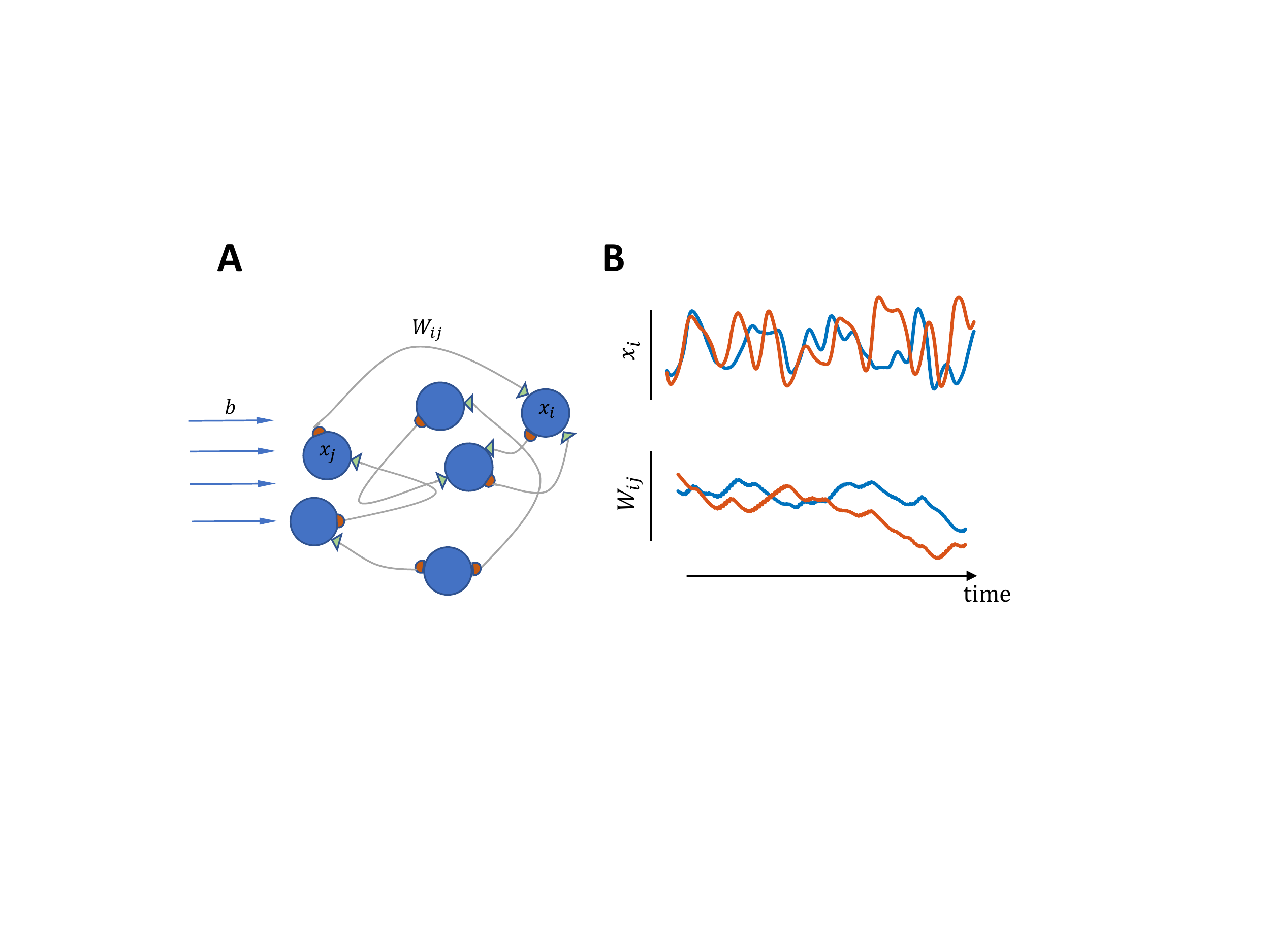}
	\vskip -4cm
	\caption{
		\textbf{Co-evolution of neural activity and connectivity.} (A) Illustration of our modeling framework: a recurrently connected neural network, with $W_{ij}$ denoting the connection strength from neuron $j$ %$\phi\left(x_j\right)$, 
		to neuron $i$. The dynamic variables $x_i$ evolve by Eq. \eqref{eq:xdot} and 
		connection strengths $W_{ij}$ evolve by Eq. \eqref{eq:Wdot}.
		An external signal $b_i$ can be added as additional input to each neuron $i$.
		(B) Both the neural state ($x_i$) and the connection strengths ($W_{ij}$) evolve over time, though on different timescales, and their dynamics are coupled.
		\label{fig:1}
	}
\end{figure}

Connectivity of task-performing networks is often designed to achieve the desired functionality, and assumed to be constant while the network is performing the task \cite{Hopfield82,Ben-Yishai1995}. 
There are models in which connectivity co-evolves with neural dynamics, but changes are usually confined to a training phase, whereas connectivity is kept constant during the test phase \cite{Sussillo2009,Kurikawa2013}. 
These models are consistent with the expectation of synaptic tenacity in the absence of learning.
In our model, to incorporate the recent observations on synaptic fluctuations, the connectivity matrix $\mathbf{W}$ continuously co-evolves with neural activity $\mathbf{x}$ throughout all task phases, albeit with a slower timescale (Fig. \ref{fig:1}, see also \cite{WeiRehearsal2014}).

%Rate models with fixed connectivity have been investigated extensively [REF?]. 
%Typically, such models assume that connectivity is fixed; 
%They are also 
%when 
%used for studying functionality and learning; then, dynamics of neural activity and connectivity are assumed to unfold over separate timescales \cite{Hopfield82,Osan11} (but see also \cite{WeiRehearsal2014,MagnascoSelfTuned2009}).
%Similarly, in our model, the connectivity matrix $\mathbf{W}$ co-evolves with neural activity $\mathbf{x}$, albeit with a slower timescale (Fig. \ref{fig:1}).

In order to study the coexistence of memory with synaptic fluctuations, we let $\mathbf{W}$ evolve due to contributions arising from both learning-related and fluctuation-related terms, denoted by $\Delta_L$ and $\Delta_F$ respectively:
\begin{equation}
	\dot{\mathbf{W}} = \eta \left(\Delta_L + \Delta_F\right), \label{eq:Wdot}
\end{equation}
with $\eta > 0$ the plasticity rate (relative to neural dynamics).
The fluctuation term includes stochastic, activity-independent noise in synaptic strength, as well as a homeostatic mechanism to control synaptic and firing-rate stability. 
These are precisely the processes which endanger the stability of an acquired memory that is assumed to be stored in synaptic patterns. 
We first consider how an existing memory is eroded by these processes, and later address the learning part and the interplay between the two.
%We first consider network-level properties of known homeostatic mechanisms, and their interplay with potential memory representations. 
%Following this analysis, we define the learning-related term and show how it gives rise to lasting memories.

\paragraph{Fluctuations erode real-coded information.}

\begin{figure}[H]	
	
	\includegraphics[width=\textwidth]{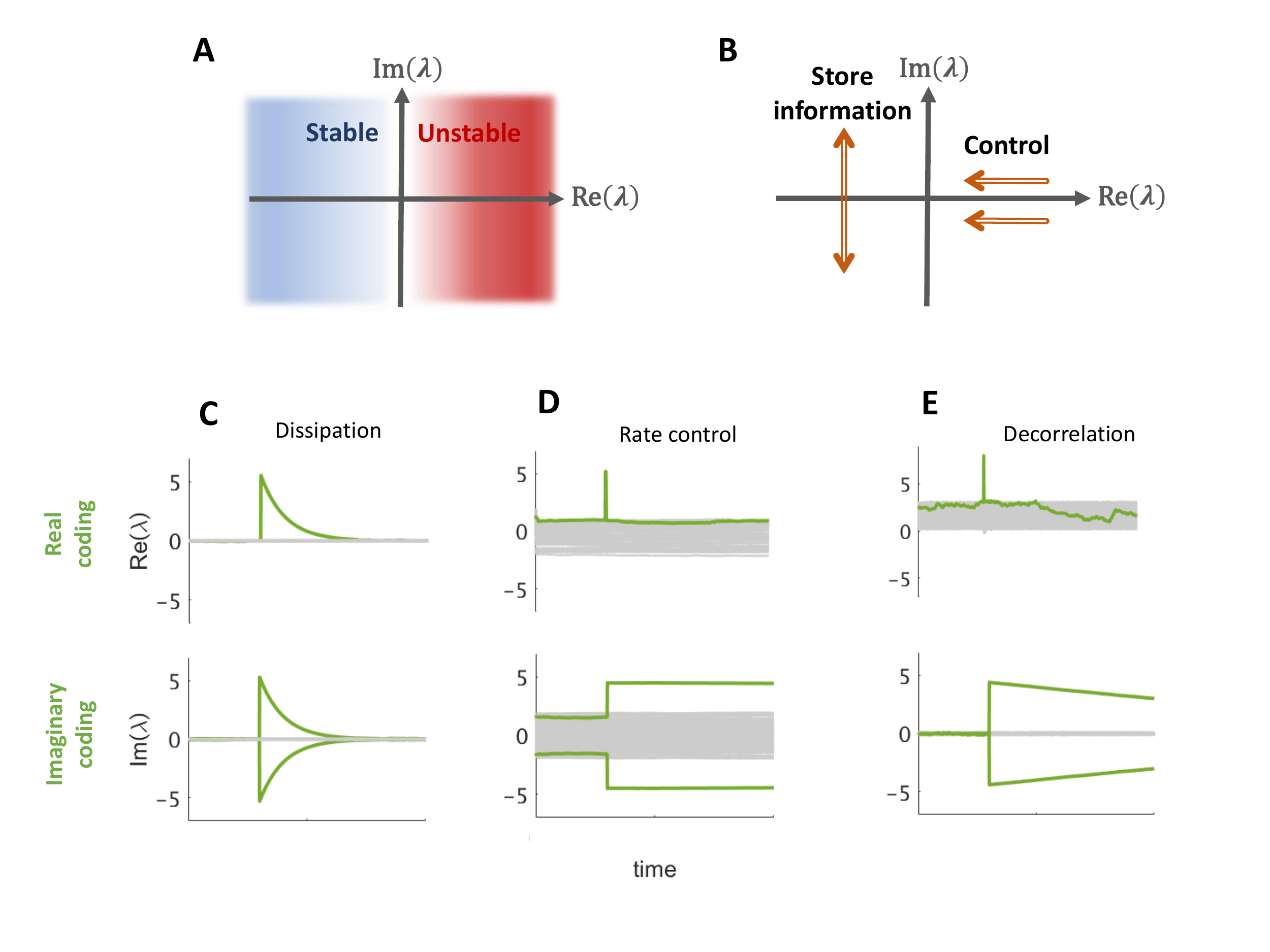}
%	\vskip -3cm
	\caption{
		\textbf{Stability and memory-associated connectivity eigenvalues}. (A) Eigenvalues of the Jacobian matrix occupy the complex plane. System stability is ensured if all eigenvalues have negative real parts (i.e. reside in the blue half-plane). Eigenvalues with positive real parts (in the red half-plane) cause divergence of trajectories with time. (B) Homeostasis mechanisms that prevent noise from accumulating and causing divergence, also affect stored memories represented by the spectrum. However, such mechanisms must only control the positive real parts, pushing them to be negative (arrows), while in the left half-plane large-amplitude imaginary eigenvalues can persist. (C,D,E) Real (top) and imaginary (bottom) parts of connectivity ($\mathbf{W}$) eigenvalue spectra for systems evolving under the dynamics of Eqs. \eqref{eq:xdot} and \eqref{eq:Wdot}, with different homeostatic mechanisms in $\Delta_F$ ($N=128$, see Methods for simulation details). 
		For each case, a memory item was embedded in $\mathbf{W}$, inducing a real outlier eigenvalue (top) or an imaginary pair (bottom).
		(C) Dissipation of synapses. Both real and imaginary memories decay with the same rate. 
		(B) Homeostatic rate-control.
		(D) Decorrelation homeostasis. In the last two, real-coded memory (top) decays rapidly whereas imaginary-coded (bottom) persists.
		\label{fig:2}
	}
\end{figure}

We model spontaneous activity-independent synaptic fluctuations by a white noise process $\xi_{ij}$ driving each synapse $ij$ independently.
Without a restraining mechanism, such dynamics would lead to divergence of the synaptic weights $W_{ij}$.
But even if the fluctuations of individual synapses were somehow bounded, this would not necessarily stabilize neural firing rates; constraining the latter requires control over network-level properties.
The stability of a dynamical system about a set-point is determined by the spectrum of the appropriate Jacobian matrix (which is the local linear approximation of the dynamics). 
The eigenvalues making up this spectrum are a collection of points in the complex plane. 
In general, the real part of this spectrum defines the system's stability: a system is only stable if all its eigenvalues have negative real parts. 
The imaginary part of the spectrum, in contrast, determines the typical timescales of small-amplitude dynamics around this set-point, but not stability itself (Fig. \ref{fig:2}A). 
Therefore, while the real part of the spectrum must be under the control of homeostatic plasticity, its imaginary part is not constrained by the requirement of stability, and is free to store information (Fig. \ref{fig:2}B). 

The arguments above derive from a general intuition on system stability; they are not a mathematical proof, as they depend on the existence of a set-point and its exact properties. 
They do, however, provide motivation to test this idea using various homeostatic mechanisms. 
We perform such tests using the connectivity matrix $\mathbf{W}$ as a proxy for the Jacobian. 
In the case of a linear network, or of linearizing around the origin, the two are equivalent. Our results below indicate that such an approximation is useful also in more general cases.

%For each proposed rule $\Delta_F$, we inject activity independent fluctuations $\xi_{ij}$ that are balanced by homeostasis. 
%Once the system has reached steady-state, we insert a memory to the network by 
Memory items are often represented in learning theory as low-dimensional perturbations to the connectivity matrix $\mathbf{W}$.
For example, in the Hopfield model a memory is associated with a particular pattern of activity $\mathbf{u}$, and is embedded in connectivity by adding a projection operator onto that pattern (of the form $\mathbf{u} \mathbf{u}^T$). 
Such a structure adds a real eigenvalue to the spectrum of $\mathbf{W}$. 
One could, however, embed different structures to $\mathbf{W}$, that add %either a real outlier eigenvalue, or 
an imaginary conjugate pair of eigenvalues. 
This defines a different type of memory item. If the above general arguments on system stability are correct, such memory items should be more resistant to synaptic fluctuations.
We test this by comparing the erosion of the two types of memory items under different homeostatic mechanisms. We first embed memories corresponding to either real or imaginary eigenvalues into the connectivity matrix $\mathbf{W}$, and then follow the dynamics of Eqs. (\ref{eq:xdot},\ref{eq:Wdot}) without active learning  ($\Delta_L=0$), 
but with various homeostatic models in $\Delta_F$.  %($\dot{\mathbf{W}}=\eta\Delta_F$). 

Perhaps the simplest implementation of a homeostatic mechanism is by dissipative synaptic dynamics, 
\[
\Delta_F = \xi - \beta \mathbf{W},   
\]
with $\beta > 0$ the rate of dissipation. 
Fig. \ref{fig:2}C shows the eigenvalues of the connectivity matrix as a function of time (gray lines), with the eigenvalues corresponding to the memory %added at $t=30$ 
highlighted in green. 
It is seen that the memory representation rapidly decays for both real (top) and imaginary (bottom) eigenvalues.
This is expected from a dissipative system, where all information decays exponentially with a rate $\beta$. 
Therefore, in the presence of such a mechanism, neither type of memory items can be sustained for longer than the decay time $1/\beta$. However, as will be shown below, this is not the case for more indirect homeostasis mechanisms.
%In reality, while synapses do have a finite lifetime and ultimately decay, this process is likely to be extremely slow and therefore cannot control synaptic fluctuation observed in experiments.

A biologically plausible homeostasis mechanism can be modelled as an activity dependent rule - where the synaptic matrix is modified to achieve a stable post-synaptic firing-rate \cite{Buonomano2005,TurrigianoSelfTuning2008,El-boustani2018}: 
\[
\Delta_F = \xi + \left(\phi_0 - \phi(\mathbf{x})\right)\phi(\mathbf{x}^T)\circ \mathbf{W},
\]
with $\phi_0$ an arbitrary target-rate vector, and $\circ$ denotes a Hadamard (element-wise) product. 
Stabilizing firing-rates around the set-point $\phi_0$ requires control over the real part of the relevant Jacobian.
Accordingly, Fig. \ref{fig:2}D (top) shows that memories stored as real eigenvalues of $\mathbf{W}$ rapidly decay. 
Imaginary-coded memories, on the other hand, may persist indefinitely without interfering with homeostasis (Fig. \ref{fig:2}D, bottom).

Finally, inspired by Ref. \cite{MagnascoSelfTuned2009}, we consider a homeostasis mechanism that does not have a well-defined firing-rate set-point. 
Instead, this rule contains an anti-Hebbian term that decorrelates firing rates across the network:
\[
\Delta_F = \xi + \mathbf{I} - \phi_\text{post}(\mathbf{x})\phi_\text{pre}(\mathbf{x}^T),
\]
where  $\phi_\text{pre}, \phi_\text{post}$ are two sigmoidal functions and $\mathbf{I}$ is the identity matrix. 
Here both connectivity and firing rates exhibit indefinite but constrained fluctuations, with
%Even though $W$ does not appear explicitly in this rule, 
the unstable modes of $\mathbf{W}$ first dominating the activity $\bf{x}$, and then being repressed by the anti-Hebbian term.

Once again, we find that the decay of imaginary-coded memories is orders of magnitude slower than that of real-coded ones (Fig. \ref{fig:2}E). 
Note that if the sigmoidal functions are identical, $\phi_\text{pre} = \phi_\text{post}$, this rule can only modify the symmetric part of $\mathbf{W}$. 
In practice, for many non-identical choices of these functions, the modification is still mostly symmetric. 
Nevertheless, the relative decay of imaginary- and real- based memories is similar to the case of the rate-control rule, that does not have any symmetric tendency.

In light of these results, a natural question arises: can a dynamical learning rule utilize the imaginary subspace to robustly code and store memory representations?

\paragraph{STDP stores imaginary-coded information.}

Symmetric and anti-symmetric matrices give rise to real and imaginary eigenvalues respectively. 
It is thus reasonable that an anti-symmetric modification to the synaptic weight matrix $\mathbf{W}$ would primarily lead to changes in the imaginary part of its spectrum. 
Local learning rules observed in experiments (e.g., STDP) have a well-defined directionality: consecutive firing of neuron $j$ before $i$ leads to a strengthening of the connection $W_{ij}$ and to the weakening of the reverse connection.  
The temporal asymmetry of STDP \cite{PooSTDP98} leads to an approximately anti-symmetric learning rule when applied to our rate model (see Methods); as such, this rule mostly affects the imaginary part of the spectrum.
In the case of perfect anti-symmetry, we find the form $\Delta_L = \phi \mathbf{y}^T - \mathbf{y}\phi^T $ ($\mathbf{y}$ is a smoothed version of $\phi\left(\mathbf{x}\right)$, see Methods), which in turn modifies only the anti-symmetric component of $\mathbf{W}$.

These arguments suggest that a biologically motivated learning rule naturally stores imaginary-coded information, thereby rendering it relatively resistant to the effect of homeostatically controlled synaptic fluctuations.
We will next investigate how such a memory can be acquired, retained and retrieved in the presence of synaptic fluctuations.
For simplicity, we will use a purely anti-symmetric $\Delta_L$.

The acquisition, i.e. the encoding and storage of a new memory trace, is initiated by stimulating the network with an external signal, $\mathbf{b}\left(t\right)$. 
A matrix with imaginary eigenvalues is necessarily of (at least) rank 2, corresponding to a two-dimensional space spanning the memory representation.
We therefore present the network with a randomly time-varying input evolving on a plane spanned by two arbitrary directions $\mathbf{u},\mathbf{v} \in\mathbb{R}^N$ (see Methods). 
As the input drives neural activity $\bf{x}$ onto the $(\bf{u},\bf{v})$ plane, the activity-dependent learning operator $\Delta_L$ follows and becomes non-negligible, which in turn causes a change in connectivity.

\begin{figure}[h!]
	
	\includegraphics[width=1.1\textwidth]{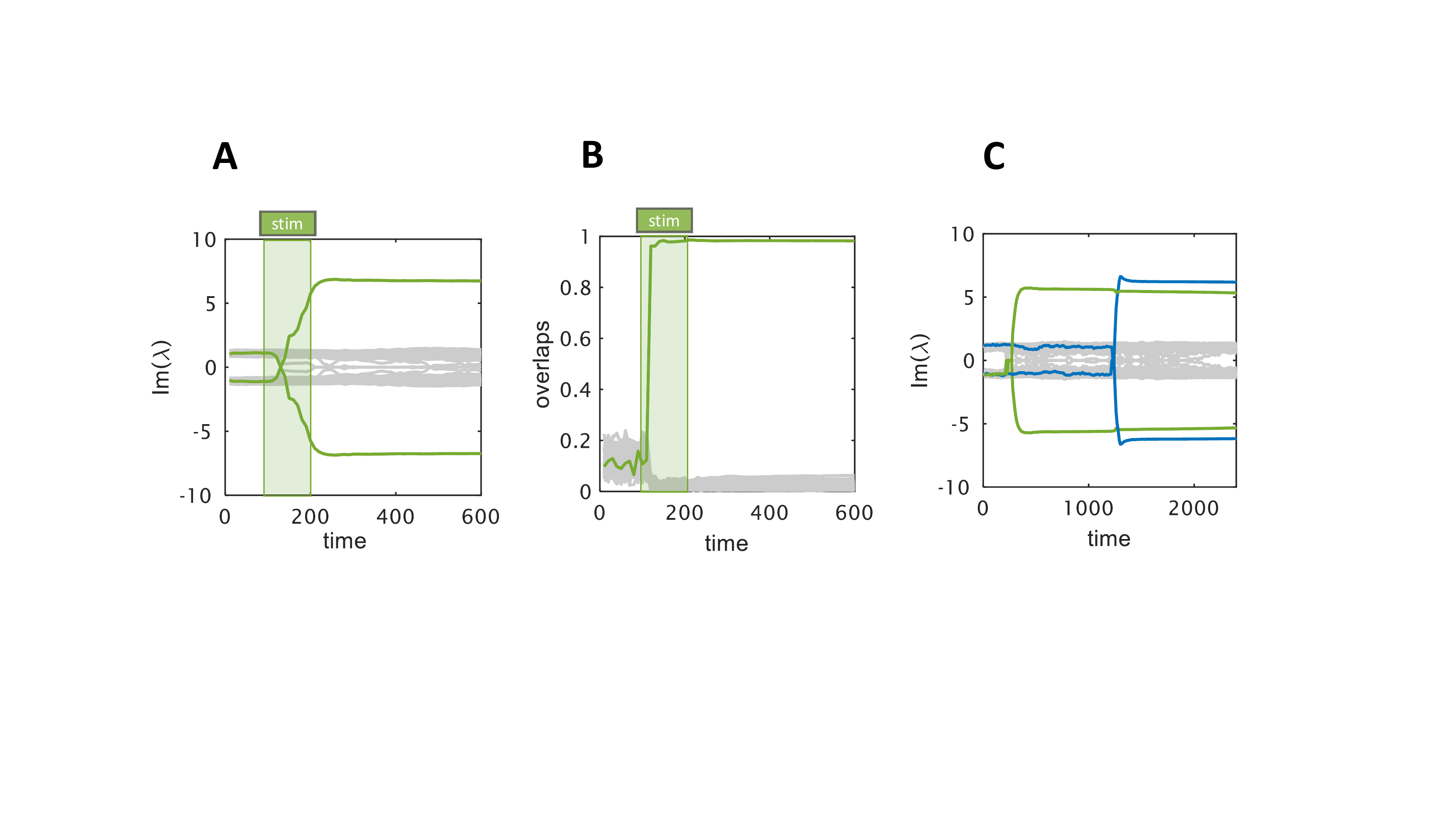}
	
	\vskip -3cm
	\caption{
		\textbf{Hebbian learning by STDP embeds persistent imaginary-coded memory.} 
		%Eigen-decomposition of the connectivity matrix $\bf{W}$ as a function of time. 
		(A) Imaginary part of the spectrum of $\bf{W}$, before, during and after external stimulation (applied between times $t=100$ and $t=200$). 
		Before learning, the imaginary part of the spectrum is almost constant in time. 
		The embedding of a memory item manifests as the growth in imaginary amplitude of one complex conjugate eigenvalue pair (green). 
		(B) During stimulus presentation, the learning rule modifies $\bf{W}$ such that the plane spanned by $\bf{u,v}$ is invariant. %$\mathbf{u}\pm i\mathbf{v}$ become eigenvectors. 
		Plotted are the overlaps of this eigenplane of $\bf{W}$, corresponding to the largest imaginary eigenvalue pair (green in $A$), with $N/2$ planes: the $\bf{u,v}$ (green), and $N/2 - 1$ orthogonal planes (gray); see Methods.
		(C) Zoom-out of panel A. 
		After the stimulus is removed, the memory representation persists (green). 
		A second stimulus, confined to a second plane, is similarly learned at a later time (blue); both memory items are retained. This figure shows results for the decorrelation rule ( $N=128$, see Methods); qualitatively similar results for rate-control not shown. 
		\label{fig:3} 
	}
\end{figure}

The learning procedure stores geometric information of the external stimulus, specifically the directions $\bf{u}$ and $\bf{v}$, within the anti-symmetric part of the connectivity matrix. 
In particular, the encoding is manifested as a rank-2 operator $\mathbf{uv}^T-\mathbf{vu}^T$ which is embedded into $\bf{W}$. 
To see this, we follow the spectrum of $\bf{W}$ as a function of time. 
During stimulus presentation, a complex conjugate eigenvalue pair forms (Fig. \ref{fig:3}A), with corresponding eigenvectors overlapping completely with the plane spanned by $\bf{u,v}$ (Fig. \ref{fig:3}B). 
The strength of the memory representation - corresponding to the magnitude of the imaginary eigenvalue - depends monotonically on stimulation duration and on input amplitude.
At later times additional stimuli may be stored using the same learning protocol (Fig. \ref{fig:3}C).

\paragraph{The nature of imaginary-coded memories.}

We have seen that a biologically plausible learning rule can capture the orientation in neuronal state-space of an incoming stimulus, and encode this information as a pair of imaginary eigenvalues in the network connectivity matrix.
%However, in order for these representations to implement memory, their information should be retrievable from the network dynamics. 
%A memory item is said to be retrieved if the neural state exhibits an activity pattern resembling the original stimulus. 
What is the nature of this memory in terms of network activity? 
We find that learning creates attractors in state-space, similar in fashion to those in the Hopfield model \cite{Hopfield82}. 
However, rather than fixed points, here the attractors are time-varying stable states - namely, limit cycles.
To see this most clearly, we consider a single imaginary-coded memory embedded in the network, and examine neural activation dynamics while keeping $\bf{W}$ fixed.
Following the Hopfield paradigm, we write $\bf{W}$ as:
\begin{equation}
%\mathbf{W}\left(t\right) \equiv 
\mathbf{W} = \rho\left(\mathbf{uv}^T-\mathbf{vu}^T\right), \label{eq:asymhop}
\end{equation}
where the coefficient $\rho>0$ represents the strength of the memory representation \cite{Coolen05}.

\begin{figure}[h!]	
	\includegraphics[width=1.1\textwidth]{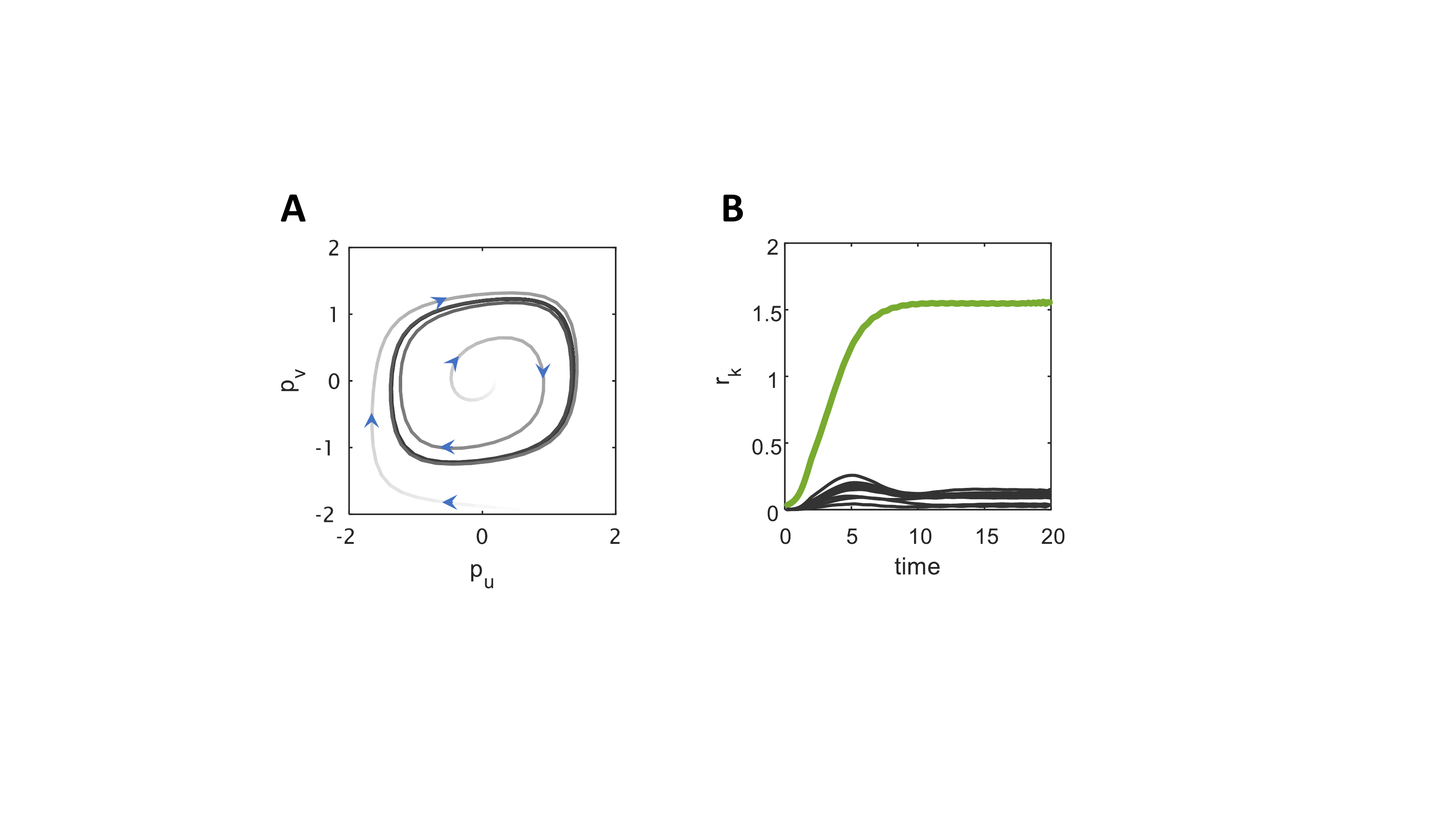}	
		\vskip -3cm
	\caption{
		\textbf{Dynamics of memory retrieval with fixed connectivity.} 
		(A) Overlaps of network activity $\bf{x}$ with the two directions spanning one embedded memory plane, $p_{\bf{u}} := \frac{1}{\sqrt{N}} \mathbf{u}^T \bf{x}$ and $p_{\bf{v}}:=\frac{1}{\sqrt{N}} \mathbf{v}^T \bf{x}$ form a stable limit cycle. 
		Shown are two trajectories, one initiated inside and the other outside the stable orbit, with arrowheads showing the direction of temporal evolution ($N=4096, ~ \rho = 4)$.
		(B) Radial coordinates of overlaps with each of multiple ($M=10,N=4096$) embedded memory planes, $r_k := \sqrt{p_{\mathbf{u}^{\left(k\right)}}^2 + p_{\mathbf{v}^{\left(k\right)}}^2}$. 
		Initiating the network near one plane, $\mathbf{u}^{\left(1\right)},\mathbf{v}^{\left(1\right)}$, results in convergence to the associated attractor (green).
		In this figure we construct connectivity by adding to Eq. \eqref{eq:asymhop} a real-coded term (see Methods).		
		\label{fig:4}
	}
\end{figure}

With one stored memory as in Eq. \eqref{eq:asymhop}, we find that, from any non-zero initial condition, the dynamics converge to periodic motion concentrated on the ‘memory plane’ spanned by $\bf{u}$ and $\bf{v}$. 
Fig. \ref{fig:4}A depicts the projections of neural activity on this plane, for two initial conditions (light gray trajectories), both converging to the limit-cycle attractor (dark closed trajectory). 
%The projections of neural activity onto the attractor plane is much larger than onto other arbitrary planes (Fig. \ref{fig:4}B). 
An approximate low-dimensional description of this limit cycle can be obtained in the limit of an infinitely steep nonlinearity $\phi$ (i.e. a step-function). 
The full dynamics are then well approximated by their projected coordinates on the plane, $p_{\bf{u}}$ and $p_{\bf{v}}$, and the low-dimensional system reads 
\begin{equation*}
\begin{aligned}
\dot{p}_{\bf{u}} &= -p_{\bf{u}} + \rho ~ q_{\bf{v}} \\
\dot{p}_{\bf{v}} &= -p_{\bf{v}} - \rho ~ q_{\bf{u}},
\end{aligned}
\end{equation*}
where $q_{\bf{v}} \approx ~ \text{arctan}\left(\frac{p_{\bf{v}}}{|p_{\bf{u}}|}\right)$ and which exhibits a stable limit-cycle around the origin (see S1 and Methods). 
We conclude that imaginary-stored memory items correspond to dynamic attractors, with geometry defined by that of the stimulating input.
This behavior stands in contrast to the classic – symmetric – Hopfield model, where memories are represented by fixed-point attractors. 

Embedding multiple memory planes $\lbrace \mathbf{u}^{\left(k\right)},\mathbf{v}^{\left(k\right)}\rbrace_{k=1}^M$ corresponds to setting
\[
\mathbf{W = UDU}^T
\]
where the columns of $\bf{U}$ are the memory patterns (interleaved $\mathbf{u}^{\left(k\right)}$ and $\mathbf{v}^{\left(k\right)}$), and $\bf{D}$ is a $2M\times 2M$ block-diagonal matrix, with the $k$-th block  reading $\bigl( \begin{smallmatrix}0 & \rho_k\\ -\rho_k & 0\end{smallmatrix}\bigr)$.
%\[
%W = \sum_{k=1}^{M}\rho_k \left(u^{\left(k\right)}{v^{\left(k\right)}}^T - v^{\left(k\right)}{u^{\left(k\right)}}^T\right) + \gamma_k \left(u^{\left(k\right)}{u^{\left(k\right)}}^T + v^{\left(k\right)}{v^{\left(k\right)}}^T\right)
%\]
Now, a locally stable limit-cycle lies on each embedded plane, and the network functions as an 
%auto-
associative memory: initiating the dynamics within the basin of attraction of one plane - providing the network with partial information of the memory to be retrieved - leads to the recovery of the full memory item (Fig. \ref{fig:4}B). 
Similar to the Hopfield model, the memory capacity is found to be proportional to system size \cite{AGSInfinitePattern85}. Numerical simulations presented in S2 show that in fact the proportionality constant is 
%namely,  the number of memories that can be stored before associative recall is destroyed by interference, tends to a constant fraction as network size increases (see numerical simulations in S2
%Via numerical simulation, the critical memory capacity of this model is found to be 
slightly higher compared with that of the symmetric Hopfield model (when normalized by a factor of two, since each memory resides on a plane).

\paragraph{Life cycle of a memory trace.}

\begin{figure}[H]
	
	\includegraphics[width=1\textwidth]{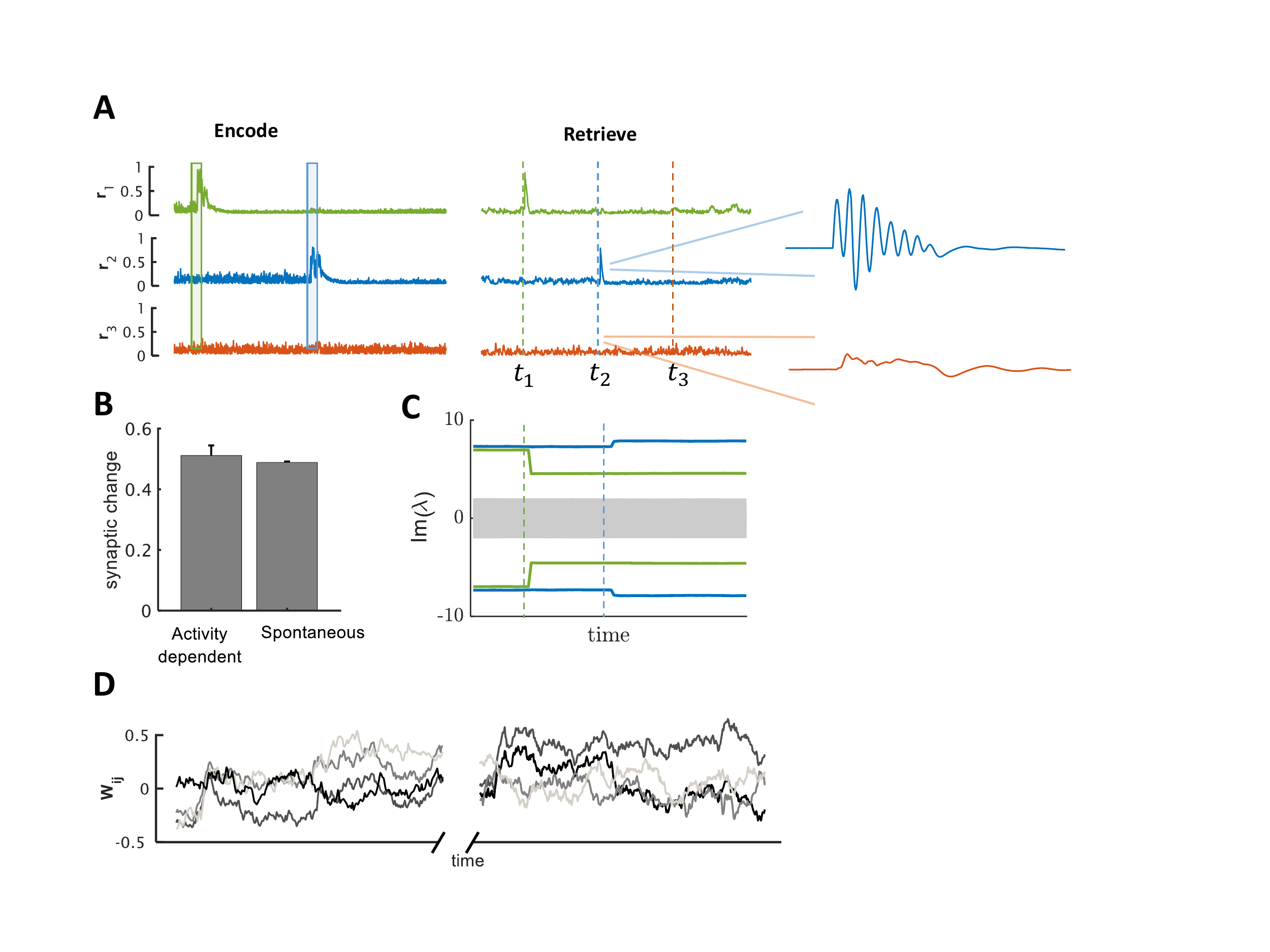}

	\caption{
		\textbf{Stable memory with unstable synapses.} (A) Left: Learning phase. 
		Two distinct stimuli are sequentially presented to the network (green, blue shades).
		Right: Retrieval phase. The embedded memories are read out of the network dynamics via a brief stimulation bearing partial information of the original stimulus. 
		Overlaps of the network state $\mathbf{x}$ with the two learned planes are shown in corresponding color. 
		During retrieval, activity follows the oscillatory trajectory of the dynamic memory item (blue zoom). 
		A novel cue, on the other hand, does not elicit a significant response in neural activity (bottom trajectory, orange).
		(B) Relative contributions of activity-dependent (left bar) and spontaneous (right bar) synaptic fluctuations are estimated to be of similar magnitude during all phases of the memory life-cycle (see Methods).
		Error bars denote one standard deviation from the mean.
		(C) Effect of retrieval on existing memories can differ, with the green memory degrading, and the blue one strengthening.
		(D) Weights of four out of $N^2$ synapses (gray shades) as a function of time. 
		The spontaneous and homeostatic contributions to plasticity drive perpetual fluctuations of synaptic weights, which occur during and after learning. 
        This figure shows results for the decorrelation rule ($N=128$, see Methods); qualitatively similar results for rate-control not shown.
		\label{fig:5}
	}
\end{figure}

We next consider the entire life-cycle of a memory in the presence of synaptic fluctuations and homeostasis, starting from learning, through retention and to retrieval.
During a learning event, implemented by presenting a stimulus in the two-dimensional memory plane, a memory representation is formed by the Hebbian learning rule. 
Fig. \ref{fig:5}A (left) shows the overlaps of neural activity onto the two planes, $r_1$ (green) and $r_2$ (blue), together with the stimulus which drives learning (shades).
These projections are elevated during stimulation, which - via the Hebbian learning rule - modifies the synaptic matrix to store each of the planes in connectivity.
The projection $r_3$ onto a third plane, which was not learned, is shown in the bottom line (orange). 

After learning, the two memory items are stored as pairs of imaginary eigenvalues, remaining stable over time, until they are retrieved at times $t_1$ and $t_2$, respectively. 
At retrieval, activity is transiently attracted to the respective memory planes, as indicated by the spikes in the overlaps (Fig. \ref{fig:5}A, right). 
During retrieval, activity follows the stored dynamic trajectory, exhibiting its typical oscillations (Fig. \ref{fig:5}A, right, blue zoom). 
At the same time, the projection onto an arbitrary plane shows no temporal structure (orange zoom). 
Finally, stimulating the network with a novel cue (at $t_3$) does not elicit a significant response in neural activity (Fig. \ref{fig:5}A; bottom trajectory, orange).

The effect of retrieval on the connectivity, namely on the stored memory itself, is somewhat unpredictable and depends on the exact state of the network and on the memory properties. 
As an example, in Fig. \ref{fig:5}C it is seen that the green memory is damaged by retrieval, namely the magnitude of the corresponding imaginary eigenvalue is decreased. 
This may be caused by the homeostatic mechanism that constrains activity, in particular the component projected onto the memory plane by the retrieval event. 
In contrast, the blue memory is slightly strengthened by retrieval, as seen by the increased magnitude of the eigenvalue pair. 
In other cases the memory remains unaffected.

Throughout this entire cycle, synapses fluctuate under the effect of activity-independent noise and homeostasis. 
Fig. \ref{fig:5}D shows a few example synapses tracked across time, during both phases. 
We may disentangle the two effects, spontaneous and activity-dependent, and estimate their relative contribution to synaptic fluctuations. 
The magnitude of the spontaneous component of synaptic fluctuations is found to be as large as that of the activity-dependent component (Fig. \ref{fig:5}B), in agreement with the experimentally observed phenomenon \cite{DvorkinRemodeling2016}.

\section*{Discussion}

Experimental evidence on the perpetual changes in synaptic strengths, with and without relation to activity or learning, has accumulated by now to form a well accepted picture: synapses are not as stable as once thought. 
Earlier theoretic work studied the statistical properties of single synapse fluctuation using phenomenological models \cite{Yasumatsu2008,LoewensteinMultiplicative2011,StatmanKestenSynapses2014}. 
These models were successful in capturing the quantitative statistics of single synapses along time and across a population, but did not address their context within an active network. 
The more difficult question of the implications of such fluctuations to network functionality has been highlighted in several recent reviews 
\cite{MongilloVolatility2017,ClopathVariance2017,ChambersReview2017,ZivTenacity17}. 

Here, we showed that it is possible to store, retain and retrieve memories in a recurrent neural network despite significant synaptic fluctuations.
Motivated from a fundamental systems-theory perspective, we argued that fluctuations, and homeostasis mechanisms that control them, place a strong constraint only on the real part of the  eigenvalues of the connectivity matrix. 
A corollary of this observation is that memories associated with imaginary eigenvalues can be kept encoded in connectivity for extended times, while synapses fluctuate under noise and homeostasis. 

We implemented this idea with the simplest form of such memories, namely two-dimensional planes in which periodic activity persists as a stable limit cycle. 
This implementation extends the classic Hopfield model, where memories are fixed-point attractors corresponding to static activity patterns, to the case where memories are represented by trajectories in neural state space \cite{Coolen05}. 
This temporal dependence is more consistent with experimental data: the oscillatory trajectories of memory-trace activations that arise naturally in our model resemble network-level oscillations observed during memory retrieval \cite{SkardaOlfactChaos87} and consolidation \cite{Pare02}. 

Since imaginary eigenvalues are associated with the anti-symmetric component of connectivity, the observed asymmetry of STDP \cite{PooSTDP2001} naturally suggests that memory items of this type can be learned dynamically. 
We have demonstrated how such learning occurs by a single stimulus presentation which spans a two-dimensional plane in activity space, thus allowing the embedding of periodic motion. 
This is most simply shown for a perfectly anti-symmetric STDP kernel, but is valid as long as the kernel contains a significant anti-symmetric component. 
The effects of symmetry of STDP on memory retention have also been noted in a different modeling context. 
In \cite{WeiRehearsal2014} the Hopfield model was studied in the presence of ongoing STDP; it was found that unstructured noise inserted into the neural state could stabilize memories with anti-symmetric, but not with symmetric, learning.

From a more general perspective, any learning rule represents the interaction of the system with its environment; if this rule is not homogeneous in space and time, the signature of this interaction might be encoded in some sub-space of network connectivity - a component with particular symmetries.
This would allow an invariant subspace of connectivity within which memories are stored, and which is minimally tampered by homeostatic fluctuations.
Indications of such invariant features have been recently observed experimentally \cite{Driscoll2017}: while individual neurons exhibited significant change in their activity patterns relative to behavior in a decision task, population activity and behavior remain stable over weeks.
The notion of an invariant subspace has also been suggested to underlie stable behavior during working memory tasks, despite ongoing neural activity \cite{Druckmann2012}.

It is also possible that other principles can be formulated that allow coexistence of stable function with synaptic fluctuations.
For example, the microscopic degeneracy of representation was shown to support stable input-output relations in a feed-forward network amid strong fluctuations \cite{Ajemian2013}.
Although the model and its implementation are very different from ours, the motivating question is broadly similar.
More recently it was proposed that in balanced cortical networks, inhibitory connectivity alone bears the burden of robust information storage \cite{Mongillo2018}, thereby rendering memories insensitive to fluctuations of excitatory synapses.

Our model was based on general considerations of system stability, without relying on specific implementations of homeostasis. Nevertheless, it gives rise to two types of experimental predictions. First, imaginary-coded memories give rise to limit-cycle attractors. Thus, retrieval of an item from long-term memory to working memory should give rise to oscillatory activity. 
These signatures of oscillations might be detected from the spectral properties of neural activity, expected to vary significantly between learning and rest phases (see S3). 
Such signatures have already been observed \cite{Osipova2006theta,Raghavachari2001} but our model suggests an additional feature that may be hiding in the data. 
The magnitude of the imaginary eigenvalue should correlate with both the oscillation frequency and with memory strength. Indeed, for a given memory item in our model, a spectral analysis correlates with memory strength. 
Unfortunately, at least in our implementation, the inter-item variability is larger than this effect (see S3).

A second prediction results from the learning rules that can give rise to imaginary-coded memories. 
Our model predicts that learning-related plasticity should be preferentially anti-symmetric. 
This could be checked by monitoring the synaptic strengths between reciprocally connected neurons during learning and rest. 
Our model predicts measurable differences in these phases (see S3). 
%This would suggest that analyzing these oscillations in more detail can provide a new

More generally, our results suggest that much systems-level understanding can be gleaned by measuring and analyzing a population of synaptic strengths across time in large networks. 
Specifically, beyond the statistical analysis of the single synapse, invariant structure in the high-dimensional space of connectivity should be searched.
Moving towards such an understanding will hopefully be possible with the advancement of experimental techniques, that will allow monitoring of multiple synapses across extended times and during various phases of behavior.

\section*{\fontsize{12}{12}\selectfont Acknowledgements}
This work was supported in part by the Israeli Science Foundation (grant number 346/16, OB; and grant number 155/18, NB). 
We thank Noam Ziv and Lukas Geyrhofer for helpful comments on an earlier version of this manuscript.

\section*{\fontsize{12}{12}\selectfont Methods}

\subsection*{\fontsize{10}{12}\selectfont \normalfont{\textit{Model simulation}}}

\fontsize{9}{12}\selectfont
We use home-made MATLAB software in order to numerically simulate Eqs. (\ref{eq:xdot},\ref{eq:Wdot}). 
The spectra of matrices are computed using built-in MATLAB functions, and their time-series sorted using the 'eigenshuffle' MATLAB script by John D'Errico, freely available online. 
For the nonlinearity in firing rates, we use the hyperbolic tangent function, $\phi\left(z\right) = \text{tanh}\left(z\right)$.
We have verified that the results presented in Fig. \ref{fig:2} are reproduced also with a rectified-linear input-output function, $\phi(z) = \text{max}(-5,z)$.
For all figures we simulate a network with $N=128$ neurons, except for the anti-symmetric Hopfield model in Fig. \ref{fig:4}B where we use $N=4096$. 
Simulations with larger networks similarly exhibit all of the discussed phenomena. 
For numerical integration we use a time constant $dt = 0.1$. 
Synaptic weights were evolved with a plasticity rate $\eta = 0.01$, this includes learning and homeostatic plasticity rules. 
For the low-pass filter $\bf{y}$, we use a first-order filter with timescale $\tau = 50$ (see next Methods section). 
For the learning process we use the time-dependent input  $\mathbf{b}\left(t\right)=c_{\bf{u}}\left(t\right) \mathbf{u} + c_{\bf{v}}\left(t\right)\bf{v}$ with ${u}_i, {v}_i \sim \mathcal{N}\left(0,\frac{1}{N}\right)$ and independent. 
The time-dependent functions $c_{\mathbf{u}}\left(t\right)$ and $c_{\mathbf{v}}\left(t\right)$ each follow an independent Ornstein-Uhlenbeck process with timescale $0.01$.
For retrieval, a brief (2 simulation time constants) pulse in the direction of $\bf{u}$ is applied to the network, namely $c_{\bf{u}}\left(t\right) = 10$ and $c_{\bf{v}}\left(t\right) = 0$.

For the rate-control homeostasis rule, we draw each component of the target-rate vector $\phi_0$ independently from a uniform distribution over the interval $\left[-1,1\right]$.
For the decorrelation homeostasis rule, we use $\phi_\text{pre} = \phi$ and $\phi_{\text{post}} \left(\mathbf{x}\right) = \text{tanh}\left(\mathbf{x} - \Bar{\mathbf{x}}\right)$, where $\Bar{\mathbf{x}}$ is a first-order low-passed version of $\bf{x}$, with timescale $\tau_x = 20$.
In all cases, we model spontaneous fluctuations by a white noise process, $\xi_{ij}\left(t\right) \sim \mathcal{N}\left(0,\frac{1}{N}\right)$, independent across time and synapses $ij$.
For the dissipative synaptic dynamics we use $\beta = 0.1$. 

In Fig. \ref{fig:3}B, each overlap is computed as the root-mean-square of the radial overlaps onto a given plane, of the two spanning directions of a second plane.

In Fig. \ref{fig:4}, we slightly modified the imaginary-coded memory representation Eq. \eqref{eq:asymhop}.
In particular, we set connectivity to $\mathbf{W} = \rho \left( \mathbf{uv}^T-\mathbf{vu}^T\right) + \gamma \left( \mathbf{uu}^T + \mathbf{vv}^T\right)$, with $\gamma > 1$.
The second term emerges naturally in $\mathbf{W}$ when the memory is learned via our dynamic learning protocol; without it, the origin in phase-space 
$\mathbb{R}^N$ becomes a locally stable fixed point and trajectories decay.
Numerically, in the limit of small integration step $dt \to 0$, we find that, for $\gamma=0$, the origin is actually globally stable, and the memory-related limit-cycle disappears.
On the other hand for discrete-time dynamics, the model with $\gamma=0$ is stable, and this is the version used for the capacity calculations.

For generating Fig. \ref{fig:5}B we compute the contribution of each plasticity term ${\Delta}(t)$ as the temporal average of $\frac{1}{N^2}\sum_{ij}| \Delta_{ij}(t) |$, from a simulation of our model with the decorrelation homeostasis rule.

\subsection*{\fontsize{10}{12}\selectfont \normalfont{\textit{Derivation of the Hebbian learning rule}}}

In this section we derive the rate-based learning rule $\Delta_L$.
Our starting point is a Poisson spiking neuron with output spiking activity $S_i\left(t\right)$ and instantaneous firing rate $\phi_i\left(t\right)$ \cite{Kempter1999}.
STDP learning is characterized by a differential update of the synaptic efficacy $W_{ij}$, based on the temporal distance $\Delta t$ between spiking of unit $i$ and unit $j$; the amplitude of change is given by the ‘learning window’ $K\left(\Delta t\right)$ \cite{PooSTDP98}. 
Denote the correlation between inbound and outbound spiking activity by
\[C_{ij} \left(t;t+\Delta t\right) = \overline{\left< S_i \left(t\right) S_j \left(t+\Delta t\right)\right> }, 
\]
where angular brackets denote ensemble averaging over the noise in spiking activity and overbar denotes temporal averaging. 
So, STDP learning can be formalized as
\begin{equation*}
\begin{aligned}
\left[\Delta_L \right]_{ij}  = \eta \int_{-\infty}^t ds K\left(t-s\right) C_{ij} \left(t;s\right) \\
+ \eta \int_{-\infty}^t ds K\left(s-t\right) C_{ij} \left(s;t\right) 
\end{aligned}
\end{equation*}
\cite{WeiRehearsal2014}.
To proceed, we approximate the correlation, writing it in terms of the instantaneous firing rates $C_{ij} \left(t;s\right) \approx \phi_i \left(t\right) \phi_j \left(s\right)$ \cite{Kempter1999}, and assume a learning window of the form 
\[ K\left(\Delta t \right) = \begin{cases} 
a_P e^{-\tau_P \Delta t} & \Delta t > 0 \\
a_D e^{\tau_D \Delta t} & \Delta t \le 0 \\
\end{cases}
\]
where $a_P,\tau_P,\tau_D>0$ and $a_D<0$. 
Performing the integration, we obtain
\[
\left[\Delta_L \right]_{ij} = \eta\left(a_P\phi_i\left(t\right) y^P_j\left(t\right) - a_D \phi_j\left(t\right) y^D_i\left(t\right)\right).
\]
where $y^P,y^D$ are first-order low-pass filters of spiking rates $\phi$; each filter is characterized by a different timescale, $\tau_P$ and $\tau_D$ respectively.

In general, the parameters of $K$ give rise to an asymmetric learning operator $\Delta_L$.
The extent of asymmetry is determined by the discrepancy between the two pairs of kernel parameters, i.e. the difference in timescales of potentiation and depression $\tau_P,\tau_D$, and the two amplitudes $a^P,a^D$. 
When $\tau_D = \tau_P$ and $a_D = -a_P$, the resulting learning operator is purely anti-symmetric:
\[
\left[\Delta_L \right]_{ij}=\eta a_P\left(\phi_i y_j - \phi_j y_i \right).
\]

\subsection*{\fontsize{10}{12}\selectfont \normalfont{\textit{Code availability}}}
Example code for simulating our main results can be found at \url{https://github.com/lsusman/stable-memory}.

\bibliography{membib}
\bibliographystyle{unsrt}

%%%%%%%%%%%%%%%%%%%%%%%%%%%%%%%%%%%%%%%%%%%%%%%%%%%%%%%%%%%%
%%%%%%%%%%%%%%%%% SUPP %%%%%%%%%%%%%%%%%%%%%%%%%%%%%%%%%%%%%

\setcounter{figure}{0}
\setcounter{equation}{0}
%\author{…} 
\renewcommand{\thesection}{S\arabic{section}}
\renewcommand{\figurename}{Figure S}
\def\fnum@figure{\figurename\thefigure}
%\counterwithin*{figure}{section}

	\titleformat{\section}
	{\normalfont\large\bfseries}{\thesection}{1em}{}

	\title{\Large{\textbf {Stable memory with unstable synapses: \newline Supplementary information}}}
	\date{}
	%\posttitle{\par\end{center}}
	
	\maketitle
	%\noindent\makebox[\linewidth]{\rule{\paperwidth}{0.4pt}}
	\vskip -2.2cm
	
	\fontsize{12}{10}\selectfont
	
	\section{Retrieval dynamics: a low-dimensional approximation}
	\label{sec:lowD}
	In the main text, neural dynamics during retrieval of an embedded memory were shown to converge to a limit-cycle attractor. 
	Here we show that these dynamics can be well approximated by a two-dimensional system.
	
	Consider the neural dynamics presented in Eq. (1) in the main text:
	\begin{equation}
	\dot{\mathbf{x}} = -\mathbf{x} + \mathbf{W}\phi\left(\mathbf{x}\right), \quad \mathbf{x}(t=0) \neq 0 \label{eq:dynamics}
	\end{equation}
	where one memory plane is embedded into the connectivity matrix:
	\begin{equation*}
	W = \rho \left(\mathbf{u}\mathbf{v}^T - \mathbf{v}\mathbf{u}^T\right) + \gamma\left(\mathbf{u}\mathbf{u}^T + \mathbf{vv}^T\right),
	\end{equation*}
	for  $\rho > 0, ~\gamma > 1$ and $\mathbf{u}, \mathbf{v} \in \mathbb{R}^N$ with independently drawn, Normally distributed components; the vectors are scaled to have unit norm.
	%For $\rho > 0$, linear stability theory predicts that the origin of $\mathbb{R}^N$ is an unstable focus. 
	Defining the projected coordinates
	\[
	p_{\mathbf{u}} := \mathbf{u}^T\mathbf{x} / \sqrt{N}, \qquad p_{\mathbf{v}} := \mathbf{v}^T\mathbf{x} / \sqrt{N},
	\]
	we have from Eq. \eqref{eq:dynamics} 
	\begin{equation}
	\begin{aligned}
	\dot{p}_{\mathbf{u}} &= -p_{\mathbf{u}} + \left(\rho \mathbf{v}^T + \gamma \mathbf{u}^T\right) \phi / \sqrt{N} \\
	\dot{p}_\mathbf{v} &= -p_{\mathbf{v}} + \left(-\rho \mathbf{u}^T + \gamma \mathbf{v}^T\right) \phi / \sqrt{N}. \label{eq:macro_dynamics}
	\end{aligned}
	\end{equation}
	%For sufficiently large $\rho$, both overlaps $p_{\mathbf{u}},p_{\mathbf{u}}$ are also large. 
	Linear stability theory predicts that the origin $0\in \mathbb{R}^N$ is an unstable focus, thus after a sufficient amount of time $\mathbf{x}$ is almost completely on the plane (all other directions are stable), $\mathbf{x} = p_{\mathbf{u}} \mathbf{u} + p_{\mathbf{v}} \mathbf{v}$.
	Now, we approximate the sigmoid $\phi$ by a step-function:
	\[
	\phi_i\left(x_i\right) = \text{sign}\left[p_{\mathbf{u}} u_{i} + p_{\mathbf{u}} v_{i}\right].
	\]
	Using this, we can approximate the coordinates of the projected rate-vector
	\begin{equation}
	\begin{aligned}
	q_{\mathbf{v}} &:= \mathbf{v}^T\phi/ \sqrt{N} \approx \sum_{i} v_{i}\text{sign}\left[p_{\mathbf{u}} u_{i} + p_{\mathbf{v}} v_{i}\right] / \sqrt{N}\\
	&= \sum_{i}|v_{i}|\text{sign}\left[p_{\mathbf{u}} \frac {u_{i}}{v_{i}} + p_{\mathbf{v}}\right] / \sqrt{N} \\
	&= \frac{1}{\sqrt{N}}\sum_{i} |v_{i}|s_i; \label{eq:qv}
	\end{aligned}
	\end{equation}
	for convenience, we have defined $s_i := \text{sign}\left[p_{\mathbf{u}} \frac {u_{i}}{v_{i}} + p_{\mathbf{v}}\right]$.
	Using this representation, we wish to express the coordinates $q_{\mathbf{u}},~q_{\mathbf{v}}$ as functions of the coordinates $p_{\mathbf{u}},~p_\mathbf{v}$. 
	First, by definition
	\[
	s_i = 1 \iff  p_{\mathbf{v}} > -\frac{u_{i}}{v_{i}} p_{\mathbf{u}}.
	\]
	Second, since $u_{i},~ v_{i}$ are independent Normal random variables, the quotient $a:=-\frac{u_{i}}{v_{i}}$ has a standard Cauchy distribution, with cumulative distribution function
	\[
	\text{Pr} \left(a \le x\right)= \frac{1}{\pi}\text{arctan}\left(x\right) + 1/2.
	\]
	Thus, 
	\begin{equation*}
	\begin{aligned}
	\text{Pr}\left(s_i = 1 \right) = \text{Pr}\left( a p_{\mathbf{u}} < p_{\mathbf{v}} \right).
	\end{aligned}
	\end{equation*}
	Assume first that $p_{\mathbf{u}} > 0$, so
	\begin{equation*}
	\begin{aligned}
	\text{Pr}\left( a p_{\mathbf{u}} < p_{\mathbf{v}} \right) &= \\
	\text{Pr}\left( a < \frac{p_{\mathbf{v}}}{p_{\mathbf{u}}} \right) &= \frac{1}{\pi}\text{arctan}\left(\frac{p_{\mathbf{v}}}{p_{\mathbf{u}}}\right) + 1/2.
	\end{aligned}
	\end{equation*}
	For $p_{\mathbf{u}} < 0$ we have
	\begin{equation*}
	\begin{aligned}
	\text{Pr}\left( a p_{\mathbf{u}} < p_{\mathbf{v}}  \right) &= \\
	\text{Pr}\left(\frac{p_{\mathbf{v}}}{p_{\mathbf{u}}} < a \right) &=
	1 - \text{Pr}\left(\frac{p_{\mathbf{v}}}{p_{\mathbf{u}}} > a \right) &=
	- \frac{1}{\pi}\text{arctan}\left(\frac{p_{\mathbf{v}}}{p_{\mathbf{u}}}\right) + 1/2.
	\end{aligned}
	\end{equation*}
	In total, we get
	\[
	\text{Pr}\left(s_i = 1 \right) = \frac{\text{sign}\left[p_{\mathbf{u}}\right]}{\pi} \text{arctan}\left(\frac{p_{\mathbf{v}}}{p_{\mathbf{u}}}\right) + 1/2.
	\]
	For $s_i = -1$ we similarly obtain
	\begin{equation*}
	\begin{aligned}
	\text{Pr}\left(s_i = -1 \right) &= 
	\text{Pr}\left(\frac{p_{\mathbf{v}}}{p_{\mathbf{u}}} < a  \right) &=
	-\frac{\text{sign}\left[p_{\mathbf{u}}\right]}{\pi} \text{arctan}\left(\frac{p_{\mathbf{v}}}{p_{\mathbf{u}}}\right) + 1/2.
	\end{aligned}
	\end{equation*}
	The expected value for $s_i$ is then
	\begin{equation*}
	\begin{aligned}
	\left<s\right>_{\mathbf{u,v}} &= \left(+1\right)\text{Pr}\left(s_i = 1 \right) + \left(-1\right)\text{Pr}\left(s_i = -1 \right) \\
	&= \frac{2}{\pi} \text{sign}\left[p_{\mathbf{u}}\right]\text{arctan}\left(\frac{p_{\mathbf{v}}}{p_{\mathbf{u}}}\right).
	\end{aligned}
	\end{equation*}
	Inserting this into the sum in Eq. \eqref{eq:qv}, approximating $|v_{i}| \approx \sqrt{\frac{2}{N\pi}}$ (the expectation of the absolute value of a Normally distributed variable), and neglecting the correlations between $|v_{i}|$ and $s_i$, we obtain
	\begin{equation*}
	\begin{aligned}
	q_{\mathbf{v}} = \frac{1}{\sqrt{N}}\sum_{i}|v_{i}| s_i &\approx \left(\frac{2}{\pi}\right)^{3/2} \text{sign}\left[p_{\mathbf{u}}\right]\text{arctan}\left(\frac{p_{\mathbf{v}}}{p_{\mathbf{u}}}\right) \\
	&= \left(\frac{2}{\pi}\right)^{3/2}\text{arctan}\left(\frac{p_{\mathbf{v}}}{|p_{\mathbf{u}}|}\right)
	\end{aligned}
	\end{equation*}
	and a similar derivation gives us
	\begin{equation*}
	\begin{aligned}
	q_{\mathbf{u}} :&= u^T\phi/\sqrt{N}  \\
	&\approx \left(\frac{2}{\pi}\right)^{3/2}\text{arctan}\left(\frac{p_{\mathbf{u}}}{|p_{\mathbf{v}}|}\right)
	\end{aligned}
	\end{equation*}
	So, inserting the expressions of $q_{\mathbf{u}}, q_{\mathbf{v}}$ into Eq. \eqref{eq:macro_dynamics}, we have
	\begin{equation}
	\begin{aligned}
	\dot{p}_{\mathbf{u}} &= -p_{\mathbf{u}} + \gamma q_{\mathbf{u}} + {\rho} q_{\mathbf{v}}\\
	\dot{p}_{\mathbf{v}} &= -p_{\mathbf{v}} + \gamma q_{\mathbf{v}} - {\rho} q_{\mathbf{u}}. \label{eq:approx}
	\end{aligned}
	\end{equation}
	
\begin{figure}[H]	
		
		\includegraphics[width=1.1\textwidth]{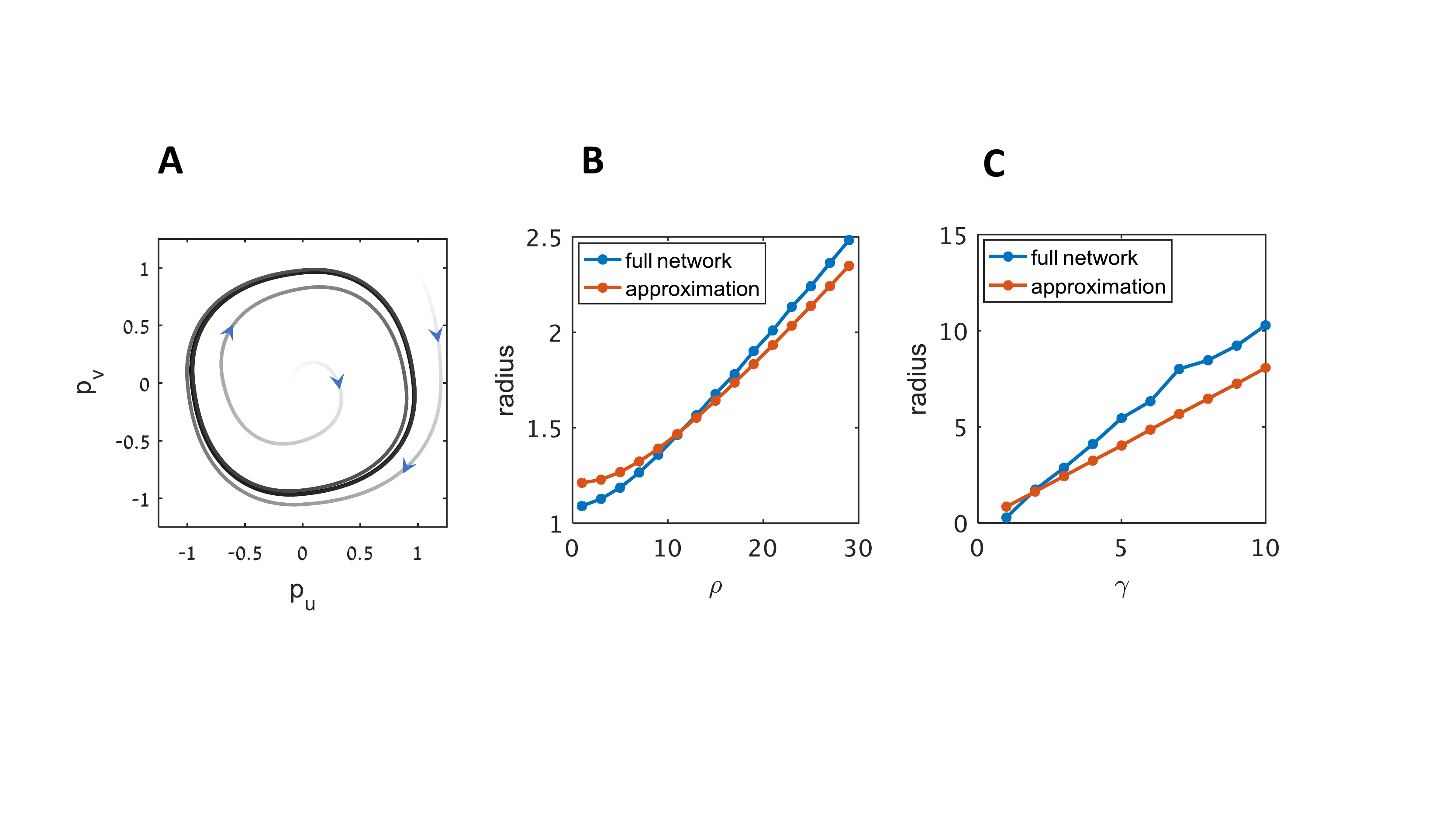}
		\vskip -2cm
		\caption{
			\textbf{Dynamics in the approximate system.} 
			(A) Simulation of the approximate system Eq. \eqref{eq:approx}, which mimics the dynamic behavior of the full system, including the radius of the limit-cycle.
			Here we use $\rho = 3$ and $\gamma = 1.5$.
			(B) Comparison of the radii of limit cycles in the full model (blue) and the approximate system (orange) for a range of $\rho$ values; $\gamma$ was kept fixed at 1.5.
			(C) Here $\rho = 3$ is fixed, and $\gamma$ varies. 
			In this case the radius varies over a greater range, and the approximation is less accurate.
			\label{fig:S2}
		}
	\end{figure}

	Numerically simulating the two-dimensional system, we indeed find a stable limit-cycle around the origin (Fig. S\ref{fig:S2}A).
	Beyond the qualitative similarity, the approximate model also captures quantitative aspects of the attractor, namely, the radius of the limit cycle over a wide range of parameter values (Fig. S\ref{fig:S2} B,C).

	\pagebreak
	\section{Capacity of the 'Limit-cycle Hopfield' model}
	
	In this section we numerically evaluate the storage capacity of our limit-cycle variant of the Hopfield model, and compare it to the classical, symmetric case \cite{Hopfield82,Hopfield84}.
	A theoretical prediction for the capacity exists only for the discrete Hopfield model \cite{Hopfield82}:
	\begin{equation}
	\begin{aligned}
	S_i\left(t+1\right) &= \text{sign} \left[\sum_{j}W_{ij}S_j\left(t\right)\right], \qquad i=1,...,N, \\ \label{eq:discrete_hopfield}
	\mathbf{W} &= \frac{1}{N}\sum_{k=1}^{M} \mathbf{u}^{\left(k\right)}{\mathbf{u}^{\left(k\right)}}^T,
	\end{aligned}
	\end{equation}
	where the dynamic variable $\mathbf{S}$, as well as the memory patterns $\mathbf{u}^{\left(k\right)}$, are binary variables. 
	%The amplitudes $\rho_k > 0$ define the combination of the $M$ different stored memories.
	
	In the limit $N, M \to \infty$, the critical memory capacity of the network is $\alpha_c = \lim_{N,M\to\infty} M/N \approx 0.14$ \cite{AGSInfinitePattern85}.
	For $\alpha > \alpha_c$, the overlap of the network state with the target pattern, $m = \mathbf{u}^T \mathbf{S} / {N}$, sharply declines from $1$.
	
	If $\mathbf{W}$ is replaced in Eq. \eqref{eq:discrete_hopfield} by any purely anti-symmetric matrix, it can be shown that the dynamics of the state $\mathbf{S}$ always converge onto a stable 4-cycle \cite{Goles86}.
	In our model, as discussed in the main text, we consider anti-symmetric connectivity matrices of the form
	\begin{equation}
	\mathbf{W} = \frac{1}{N}\sum_{k=1}^{M}\left(\mathbf{u}^{\left(k\right)}{\mathbf{v}^{\left(k\right)}}^T - \mathbf{v}^{\left(k\right)}{\mathbf{u}^{\left(k\right)}}^T\right). \label{eq:W}
	\end{equation}
	% + \gamma_k \left(u^{\left(k\right)}{u^{\left(k\right)}}^T + v^{\left(k\right)}{v^{\left(k\right)}}^T\right),
	%where $\rho_k > \gamma_k \ge 0$.
	The stable cycles arising from this connectivity are precisely $\lbrace \pm \mathbf{u}^{\left(k\right)}, \pm \mathbf{v}^{\left(k\right)} \rbrace$, thus linking the result of \cite{Goles86} to the geometry of the eigenspace of $\mathbf{W}$.
	
	When assessing the capacity of this model, one must take into account the 
	%oscillatory, 
	two-dimensional nature of the attractor states.
	Thus, we use the $L_1$ radius on each embedded plane, $m = |q_{\mathbf{u}}^2| + |q_{\mathbf{v}}^2|$ with $q_{\mathbf{u}} = \mathbf{u}^T \mathbf{S} / {N}$, and, for comparison with the symmetric model, a given $\alpha$ is computed for a network embedded with $M/2$ memory planes.
	In other words, in both cases we are counting the total dimensionality of memory-space.
	
\begin{figure}[H]	
		
		\includegraphics[width=1.2\textwidth]{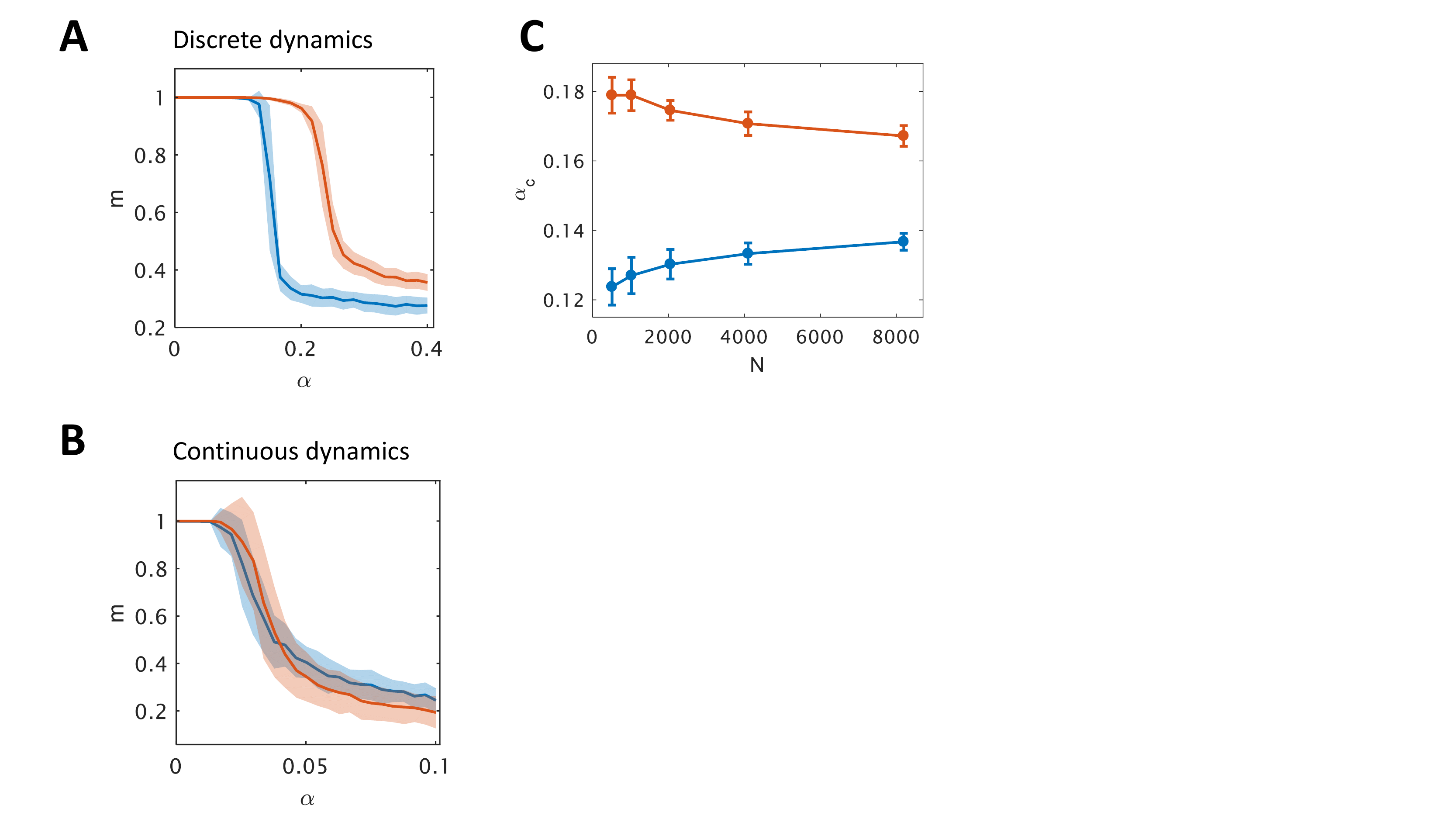}
		\caption{
			\textbf{Capacity of the Hopfield model.}
			Overlap with a target memory item as a function of memory loading $\alpha = M/N$, for the discrete (A) or continuous (B) models with $N=4096$ neurons (and $\rho = 8$ in the continuous case); the symmetric model is plotted in blue and the anti-symmetric in orange.
			Lines represent the final overlap averaged over 100 realizations (each evolved over $T=50$ time-steps), shaded areas mark standard deviation from the mean.
			(C) The critical memory load, the $\alpha$ yielding at least $2\%$ error in overlap, for a range of network sizes $N$.
			Data points represent average over 20 realizations of the computation in (A), bars mark standard deviation.
			In all cases we use random patterns with $P\left(u_i = \pm1\right) = 0.5$ independently.
			In each instance, the network is initiated at a noisy version of the target pattern.
			For the discrete model, we randomly flip $10\%$ of the bits in the desired target pattern; for the continuous case, a zero-mean, 0.2 standard deviation Gaussian variable is added to each component independently.
			\label{fig:S3}
		}
	\end{figure}

	Fig. S\ref{fig:S3}A shows the overlaps obtained from simulating the two models (Equations \eqref{eq:discrete_hopfield},\eqref{eq:W}; blue: symmetric Hopfield, and orange: our model, respectively), for a fixed $N$ and varying memory loads $\alpha$.
	The critical load is found to be slightly higher for the anti-symmetric model.
	Varying the network size $N$, we find that in both models the critical capacity approaches a constant limit - namely, the number of memories that can be embedded in the network scales with $N$ (Fig. S\ref{fig:S3}C).
	Interestingly, while the symmetric model is bounded by the theoretical prediction of $\alpha_c\approx 0.14$, the anti-symmetric variant has a higher capacity - for any finite $N$, and, presumably, asymptotically.
	
	We checked also the capacity in the continuous version of our model, and compared it to the analogous Hopfield model \cite{Hopfield84}.
	Again, the limit-cycle variant shows a slightly higher maximal memory load compared to the classical model; for both continuous variants, the capacity is reduced relative to the discrete case (Fig. S\ref{fig:S3}B).
	
	\pagebreak
	\section{Model predictions}
	
	A central concept in our model is that of imaginary-coded memory.
	This notion corresponds to properties of network activity as well as connectivity, properties which may be estimated from data recorded during learning and during rest. 
	
	Assume first that our dataset includes the temporal evolution of all synaptic weights within a subnetwork storing a memory trace.
	In our model, which assumes fully-connected networks, homeostasis spares imaginary-coded memories, and STDP learning generates such.
	We have validated numerically that, also with sparse connectivity, this remains true (not shown).
	So given such data, one could track the eigenvalues of the connectivity matrix, and our model predicts a significant increase of imaginary amplitude of a small number of eigenvalues (as presented in the main text).
	
	However, with current imaging technology, it is unrealistic to assume trajectories of all existing connections, so we discuss now the more realistic case, where only a subset of recorded weights are at hand.
	We do assume that recorded connections are bidirectional, and that both directions are tracked. 
	As an outcome of imaginary-coded memory, learning should induce strongly anti-correlated fluctuations of reciprocal connections; a homeostasis mechanism which stabilizes learning, by virtue of real-part control, is expected to drive positive correlations.
	
\begin{figure}[H]	
		
		\includegraphics[width=200mm]{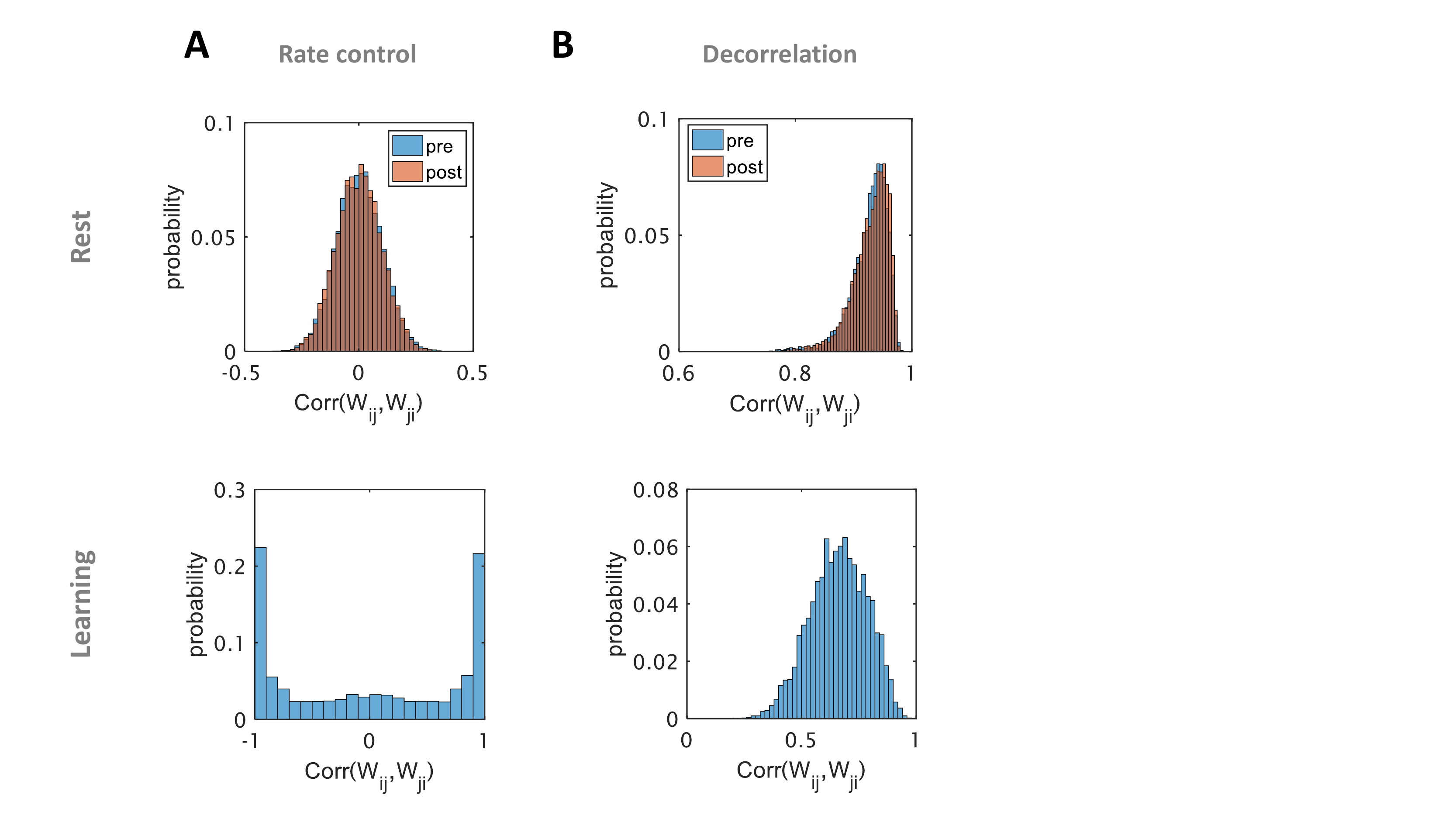}
		\caption{
			\textbf{Reciprocal fluctuations during learning and rest.}
			Histograms of correlation coefficients between changes of reciprocal connections $\dot{W}_{ij}$ and $\dot{W}_{ji}$, for the rate-control (A) and decorrelation (B) homeostasis mechanisms.
			Top: correlations at rest, before and after learning. 
			Bottom: correlations during learning.
			All correlations are computed from a network with $N=128$ neurons, all other parameters as detailed in Methods.
			\label{fig:S9}
		}
	\end{figure}

	We assess these predictions on the two homeostasis mechanisms considered.
	During rest, in the case of rate-control homeostasis, steady-state fluctuations result in a unimodal zero-mean distribution of correlations (Fig. S\ref{fig:S9}A top).
	With decorrelation homeostasis, symmetric-dominant plasticity induces correlated fluctuations (Fig. S\ref{fig:S9}B top), forming a unimodal distribution with positive mean.
	The difference between the two homeostasis mechanisms becomes more dramatic during learning.
	With rate-control, learning drives anti-symmetric synaptic change in the direction of the stimulus, whereas homeostasis acts to stabilize the set-point $\phi = \phi_0$.
	Since these activity patterns are generally unaligned, the learning and homeostasis terms have distinct eigenspaces, and the result is a bimodal distribution of correlation coefficients (Fig. S\ref{fig:S9}A bottom).
	
	The picture is qualitatively different with decorrelating homeostasis: during learning, the anti-Hebbian term strongly suppresses the same directions that are learned.
	This alignment of eigenspaces results in a cancellation between the effects of learning and homeostasis, such that overall, reciprocal connections are positively correlated (Fig. S\ref{fig:S9}B bottom).
	Nevertheless, the learning and rest phases can be quantitatively distinguished, as the distribution is shifted towards independence.
	Finally, after learning the correlations relax back to their respective steady-state distributions (Fig. S\ref{fig:S9} top, orange).
	
	These observations suggest that the existence of global subspaces underlying memory can be probed from partial data, specifically by characterizing the statistics of correlations between the fluctuations of reciprocal connections. A significant difference between learning and rest in these statistics would imply the importance of the imaginary eigenvalues in memory; moreover, the specific form of distributions might provide indirect information on homeostatic mechanisms in the network.

\begin{figure}[H]	
		
		\includegraphics[width=1.3\textwidth]{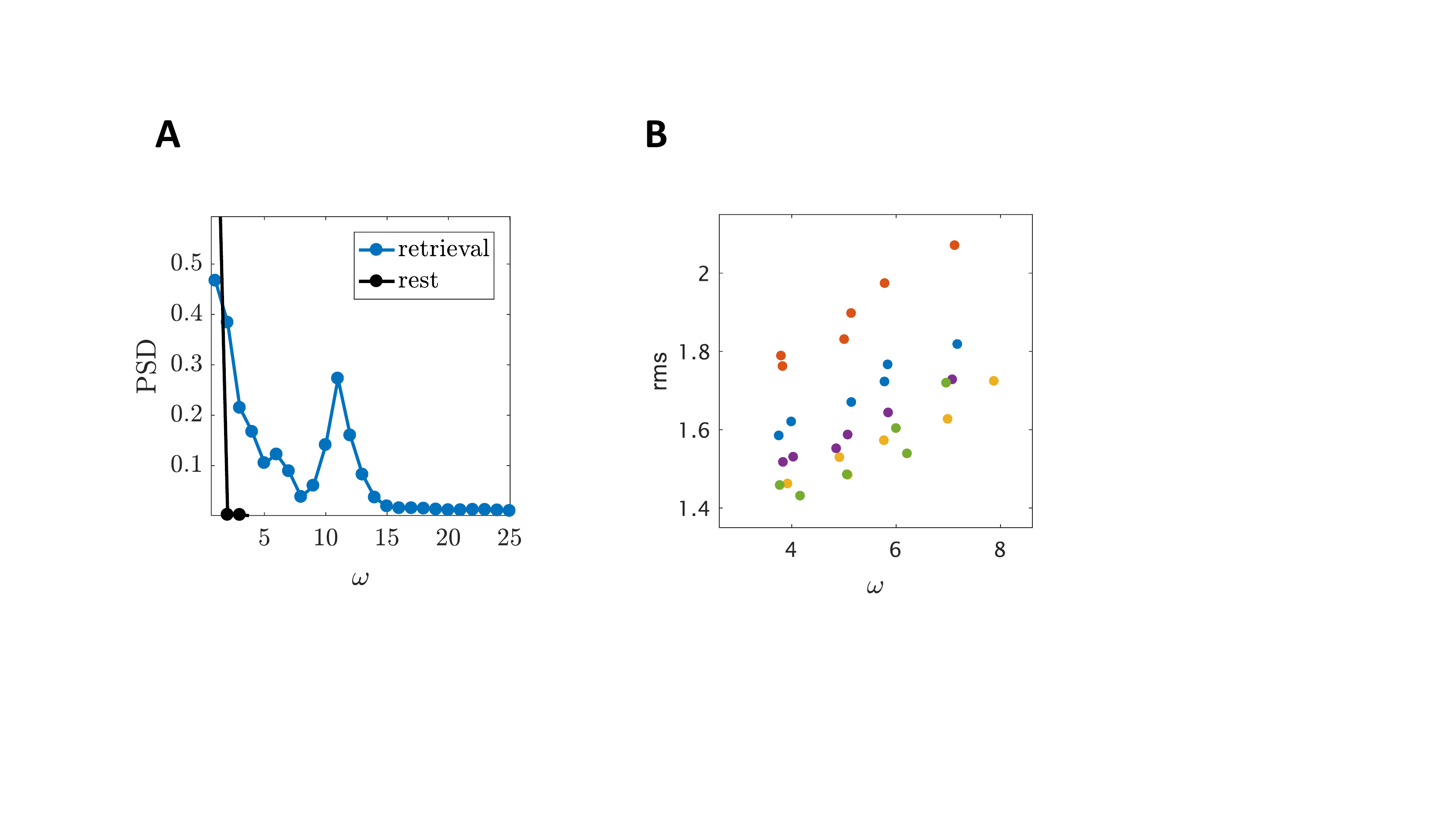}
		\caption{
			\textbf{Population-level oscillations during memory retrieval.}
			(A) Power-spectral density of the projected network activity ($p_{\bf{u}} = \frac{1}{N}\mathbf{u}^T\bf{x}$) during retrieval (blue) and during rest (black).
			(B) Root-mean-square of the projected coordinate $p_{\bf{u}}$ during 6 retrieval events of 5 different realizations of a stored memory (colors).
			In both panels we simulate our model with the rate-control homeostasis rule and $N=128$.
			\label{fig:S9-2}
		}
	\end{figure}

	Finally, our model makes statements about neural activity during memory retrieval.
	As we have seen, imaginary-coded memory implies oscillatory motion during retrieval, and furthermore, the strength of representation corresponds to the magnitude of the imaginary part of the coding eigenvalue.
	This network-level feature manifests as the dominant frequency in the power-spectral density of activity during retrieval (Fig. S\ref{fig:S9-2}A, blue).
	In contrast, no such frequency exists in the population activity at rest (Fig. S\ref{fig:S9-2}A, black).
	
	Our model also offers a relation between the strength of a memory trace and the above discussed spectral content: stronger memories correspond to higher dominant frequency during retrieval.
	Fig. S\ref{fig:S9-2}B shows a measure of the population-response salience as a function of the dominant frequency, for multiple retrieval events of a collection of memory traces (colors).
	Note that, while these two measures show a high correlation for each individual memory, the across-memory variance may be of larger extent.
	This suggests that our measure of salience is not suitable for observing the predicted relation, but we hope that in realistic experimental situations a more informative measure of memory strength may be found in behavioral performance.

\end{document}